\documentclass[final,5p,times]{elsarticle}

\usepackage{amssymb}
\usepackage{amsmath}
\usepackage{physics}
\usepackage{hyperref}
\usepackage{braket}
\usepackage{longtable}
\usepackage{multirow}
\usepackage{listings}
\usepackage{xcolor}

\newcounter{bla}

\journal{Computer Physics Communications}
\newcommand{\comment}[1]{}

\graphicspath{ {./figures/} }

\begin{document}

\begin{frontmatter}
\title{Centroids of nuclear shell-model Hamiltonians, with optimization of energy-based truncation schemes}

\author[add1]{Calvin W. Johnson}\ead{cjohnson@sdsu.edu}

\author[add1]{Austin Keller}

\address[add1]{San Diego State University, San Diego, California 92182, USA}

\begin{abstract}
The  configuration-interaction shell model is an effective and widely-used approach to the nuclear many-body problem, whose main drawback is the exponential growth of the basis dimension.  An useful way to character nuclear shell-model Hamiltonians 
is through traces, including traces in subspaces defined by orbital occupations.  Such traces, or energy centroids, can be easily and efficiently computed through the monopole components of the nuclear interaction, that is, terms that go like 
$n_a n_b$ where $n_a$ is the occupation of the  single-particle orbital labeled by $a$. These calculations can be carried out very quickly for both 
empirical (valence space) and no-core shell model spaces and interactions.
In fact, they can be carried out so fast, one can use this to optimize an 
efficient, if approximate, many-body truncation scheme used in available 
nuclear shell-model codes such as {\tt BIGSTICK}.  To carry out both 
the traces and the optimization, we present the {\tt tracer} code, 
written in Fortran90 and 
described and available here.  We give example results as well as discuss performance. 
\end{abstract}

\begin{keyword}
nuclear physics, shell model, configuration-interaction, Hamiltonian trace, monopole interaction,  energy centroid,
energy-based truncation, Fortran
\end{keyword}

\end{frontmatter}




\section{Introduction}

A widely used methodology for the nuclear many-body problem is configuration interaction (CI), usually in a spherical shell-model basis~\cite{BG77,ca05}. The configuration-interaction shell model is amenable to a wide variety of forces, can generated 
excited states and compute multiple observables such as transition  with relative ease, and is transparent not only in the underlying theory but also provides an explicit and accessible representation of the many-body wave function. The most critical drawback, however, is that the basis dimension grows exponentially. This not only spurs the use of alternate methods, such 
as coupled-clusters, which grow polynomially, but also leads one to search for effective truncation schemes for CI calculations. 

One approach to mitigating the challenges of large basis dimensions is to characterize the nuclear many-body Hamiltonian by its moments, which in turn can be expressed in terms of traces~\cite{french1967measures,PhysRevC.3.94,french1980elementary,french1983statistical,wong1986nuclear}.  Given a Hamiltonian $\hat{H}$ 
with eigenpairs $\hat{H} | \Psi_i \rangle = E_i | \Psi_i \rangle$, then the 
dimension $d$ of the Hilbert space can be written as 
\begin{equation}
    d = \sum_i 1 = \mathrm{tr} \, \mathbf{1},
\end{equation}
and the $k$th  moment is written as an average,  
\begin{equation}
    \mu^k = \frac{1}{d}\sum_i E_i^k = \frac{1}{d} \mathrm{tr} \, \hat{H}^k.
\end{equation}
The strategy of characterizing the Hamiltonian in terms of its moments is 
known as \textit{spectral distribution theory} (SDT) or \textit{nuclear statistical spectroscopy}. It can be used to, for example, approximate the density of states~\cite{ayik1974shell,mon1975statistical,PhysRevLett.51.2183,johnson2001statistical,nabi2001reliable,PhysRevC.74.067302} or provide a measure of the similarity betwen two Hamiltonian~\cite{launey2014program}, among other applications~\cite{kar2015study,PhysRevC.40.1826}.  The downside  of spectral distribution theory is that, by being an average over the bulk of the spectrum, it is less sensitive to, and consequently less accurate in constraining, the low-lying states which are in many cases the focus of experiment. Nonetheless, one can still learn significant facts from 
an SDT approach, for example how a Hamiltonian evolves under renormalization~\cite{johnson2017tracing}.

Here we focus only on the first moment, or centroid, which is the 
average energy in the space. Other papers~\cite{PhysRevC.3.94,wong1986nuclear,PhysRevC.8.135,ayik1974configuration}
and computer codes~\cite{chang1982jp,launey2014program} discuss  higher moments. 

Rather than tracing over the entire space, one can subdivide the Hilbert space into subspaces, labeled by $\alpha$. 
Consider the subspace projection operator,
\begin{equation}
    \hat{P}_\alpha = \sum_{i \in \alpha}  | i \rangle \langle i |,
\end{equation}
where we label orthonormal states by $i$, then the subspace dimension is a simple trace,
\begin{equation}
    d_\alpha = \mathrm{tr} \, \hat{P}_\alpha.
\end{equation}    
Similarly one can compute the  subspace first moment as a trace, 
$\mathrm{tr} \, \hat{P}_\alpha \hat{H}$.   More useful is 
the related subspace centroid, which is the average value of the diagonal matrix elements:
\begin{equation}
    E_\alpha = \frac{1}{d_\alpha}  \mathrm{tr} \, \hat{P}_\alpha \hat{H} 
    = \frac{1}{d_\alpha} \sum_{i \in \alpha} \langle i | \hat{H} | i \rangle. \label{eq:centroid}
\end{equation}
If one uses orbital occupations (or configurations) to partition the Hilbert space,
then, as detailed in Section \ref{sec:centroids}, the \textit{configuration} 
dimensions and centroids can be computed easily and efficiently without having to explicitly compute 
the many-body Hamiltonian matrix elements. 
(Unfortunately, centroids in other partitioning schemes are much more challenging. Even 
partitioning by angular momentum, while possible, is computationally 
 time-consuming~\cite{PhysRevC.69.041307,PhysRevC.82.024304}.)  
\begin{figure}[h]
    \centering
    \includegraphics[width=0.5\textwidth]{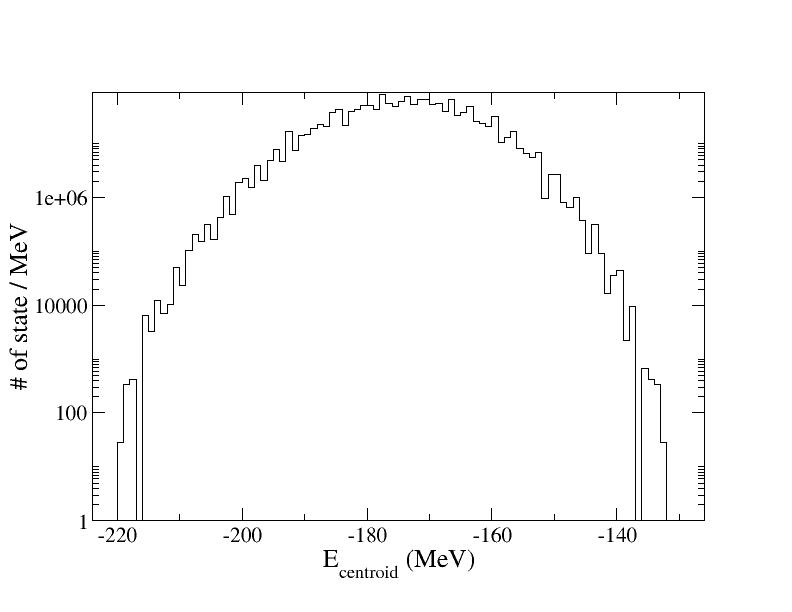}
    \caption{Distribution of number of states (not levels) in configurations versus configuration centroids for $^{60}$Fe computed in the $1p0f$ with the GX1A interaction. Data binned in 1-MeV bins.
    }
    \label{fig:fe60}
\end{figure}
\begin{figure}[h]
    \centering
    \includegraphics[width=0.5\textwidth]{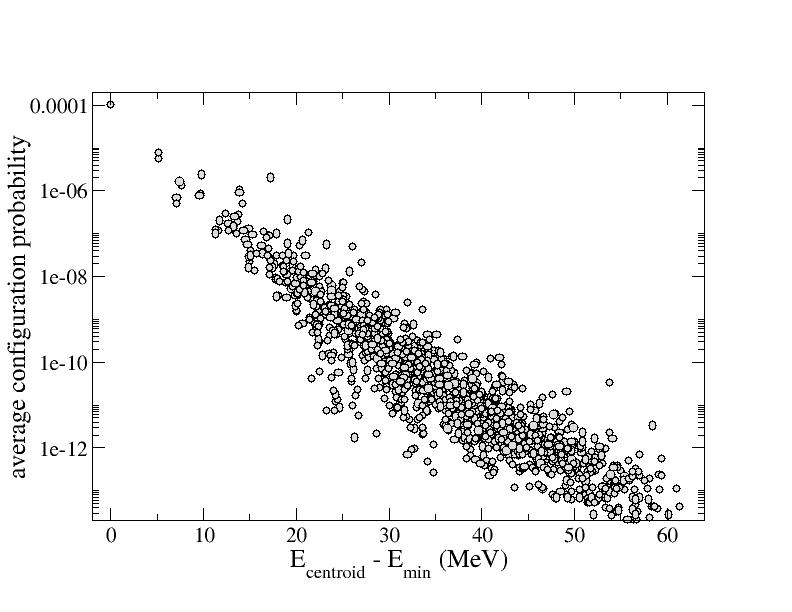}
    \caption{Average probabilities of configurations as a function of configuration energy centroid,
    for the ground state of $^{49}$Cr computed in the $1p0f$ shell using the GX1A interaction~\cite{PhysRevC.65.061301,PhysRevC.69.034335,honma2005shell}. 
    The energy centroids, Eq.~(\ref{eq:centroid}), are defined relative to the lowest centroid.
    Because each occupation configuration, e.g. $(0f_{7/2})^9$ contains multiple basis states, 
    and we give only one probability per configuration, the probabilities here do not sum to one.
    }
    \label{fig:cr49configs}
\end{figure}

One practical application of the configuration centroids is a 
truncation scheme for the shell model. One would expect low-lying eigenstates 
of $\hat{H}$ to be dominated by basis states in configurations with the 
lowest energies.  This has been suggested 
previously~\cite{PhysRevC.50.R2274}.  Fig.~\ref{fig:cr49configs} shows the probability of 
occupation configurations as a function of configuration centroid energy (relative to the 
lowest centroid), for the ground state of $^{49}$Cr in the $1p0f$ shell above a 
frozen $^{40}$Ca core,  using the interaction GX1A~\cite{PhysRevC.65.061301,PhysRevC.69.034335,honma2005shell}.
The energy centroids were calculated using {\tt tracer}, while the configuration probabilities computed using {\tt BIGSTICK}.  
While there is some scatter, 
the probabilities clearly fall off exponentially with energy.

Of course, such a truncation scheme is not perfect, due 
to the residual interaction, and collective interactions coupling 
across subspaces can lead to ``intruder'' states, that is, eigenstates which lie low in the spectrum but with significant fraction of components belonging to 
subspaces with high centroids.  One possible alternative would be to use a partitioning which addresses collectivity, for example, partitions based upon seniority or groups such as Elliott's SU(3)~\cite{dytrych2016efficacy} or the symplectic group Sp(3,R)~\cite{dytrych2008ab,launey2016symmetry,PhysRevLett.125.102505}. While this idea motivates alternative bases for the shell model, 
in practice computing subspace centroids for such group-theoretical partitions is not currently possible, and practitioners either select the basis subspace 
\textit{ad hoc}, or, more successfully, extrapolating from smaller spaces to larger spaces~\cite{PhysRevLett.82.2064}.

A truncated basis constructed along subspace partitions can be time-consuming to generate, and, for subspaces defined by non-Abelian group irreps, it can be even more time-consuming to obtain the Hamiltonian matrix elements.  
Thus we turn to orbital occupation partitions, which while imperfect are the most practical. 

Even for orbital occupation partitions, constructing a truncated basis by a sharp cutoff in the subspace energy centroid requires complicated bookkeeping. There is a simpler, if approximate, variation of this scheme, where one assigns integer weights to orbitals, and then imposes a maximum total weight.  Such a scheme lends itself to efficient algorithms~\cite{BIGSTICK}. 

The integer orbital weights can be visualized as an approximation to non-integer effective single-particle energies, which in turn approximate the differences in orbital occupation subspace centroids. As an extension of 
our configuration centroid code, we developed a relatively simple Monte Carlo algorithm to 
optimize the weights so as to approximate a cutoff in centroid energy. 
This is detailed in Sec.~\ref{sec:truncate}. 

Here we present the code {\tt tracer}, written in Fortran90 with optional OpenMP parallelization. 
Our code shares formats with the {\tt BIGSTICK} CI code~\cite{johnson2018bigstick,BigstickCode}, though it could easily interface with other CI codes. {\tt tracer} computes the occupation subspace centroids for two-body Hamiltonians (three-body Hamiltonians are not yet implemented), as described in Sec.~\ref{sec:centroids}, and, optionally, can find an approximate optimized truncation weight scheme, detailed in Sec.~\ref{sec:truncate}.
The code can be easily compiled and run, and Sec.~\ref{sec:run} gives a description 
of the input file formats and examples runs and output.

\section{Nuclear shell-model Hamiltonians}

We briefly discuss here the representation of nuclear Hamiltonians in a spherical shell-model basis~\cite{BG77,ca05}. We start from a basis of single-particle orbitals with good angular momentum, labeled 
by indices $a,b$, which themselves are shorthand for the standard orbital quantum numbers: 
orbital angular momentum $l_a$, intrinsic spin $s_a$ (which is the same for all orbitals and 
for nuclear systems is $1/2$, although the formalism is insensitive to this), total angular momentum $j_a = | l_a \pm s_a |$, and 
any radial quantum number $\nu_a$, such as the number of nodes in the radial wave function. 
Finally, each orbital has an isospin projection $t_{z,a} = +1/2$ for protons and $=-1/2$ for neutrons. 
(Single particle \textit{states}, on the other hand, have the same quantum numbers as orbitals 
but are further distinguished by the $z$-component of $j$, $m$.)

Key to our formalism are single-fermion creation and annihilation operators, 
$\hat{a}^\dagger_{a,m_a}$ and $\hat{a}_{a,m_a}$, respectively. We also use 
the pair-creation operator, 
\begin{eqnarray}
    \hat{A}^\dagger_{JM}(ab) & \equiv & \left [ \hat{a}^\dagger_a \times 
    \hat{a}^\dagger_b \right ]_{JM}  \\
    \nonumber & = & \sum_{m_a, m_b} 
    (j_a m_a, j_b m_b | JM ) \,  \hat{a}^\dagger_{a, m_a}\hat{a}^\dagger_{b, m_b}.
\end{eqnarray}
Here $(j_a m_a, j_b m_b | JM )$ is a Clebsch-Gordan coefficient~\cite{edmonds1996angular}. 
The pair destruction operator is $\hat{A}_{JM}(ab) = \left (  \hat{A}^\dagger_{JM}(ab) \right  )^\dagger$.
Finally, we let $\hat{n}_a$ be the number operator for orbital $a$. 

We now define the nuclear Hamiltonian as 
\begin{equation}
  \hat{H} =  \sum_{a} \epsilon_a \hat{n}_a + \frac{1}{4}
  \sum_{abcd} \zeta_{ab} \zeta_{cd} V_{J}(ab,cd) \sum_M \hat{A}^\dagger_{JM}(ab) 
  \hat{A}_{JM}(cd),
\end{equation}
where $\zeta_{ab} = \sqrt{1+\delta_{ab}}.$
The $\epsilon_a$ are the bare single-particle energies 
(not to be confused with the effective single-particle energies, to be discussed below);
one can generalize to off-diagonal single-particle terms, but such play no role in our formalism.
The $V_J$ are the two-body matrix elements, 
\begin{equation}
    V_J(ab,cd) = \left \langle ab, JM \left | \hat{H} \right | cd, JM \right \rangle,
\end{equation}
that is, matrix elements of the Hamiltonian between normalized 
two-body states,
\begin{equation}
    |ab, J M \rangle = \frac{1}{\sqrt{1+ \delta_{ab}}} \hat{A}^\dagger_{JM}(ab) | 0 \rangle.
\end{equation} 

This formalism applies both to valence space empirical shell model calculations and to the no-core shell model. Any and all assumptions 
and choices about, for example, the radial component of the 
single-particle orbitals, is absorbed into the values of the matrix elements.  No-core shell model calculations can include three-body forces; such an extension is possible here, but not yet implemented.

\section{Monopole terms and traces}
\label{sec:centroids}

The approach described in this paper is made practical by the realization, 
long ago, that one can compute traces of Hamiltonians in many-body spaces without 
explicitly computing the diagonal elements of the many-body matrix. Instead, one can evaluate the traces by using the part of the Hamiltonian that can be written in terms of number operators, the so-called monopole Hamiltonian~\cite{PhysRevC.3.94,wong1986nuclear,PhysRevC.59.R2347}:
\begin{equation}
    \hat{H}_m = \sum_{a} \epsilon_a \hat{n}_a + \frac{1}{2} U_{ab} \hat{n}_a
   ( \hat{n}_b -\delta_{ab} ).
\end{equation}
Again we remind the reader that here we take protons and neutrons to have 
separate orbitals.

We use subspaces $\alpha$ defined by orbital occupation: $\alpha = \{ \vec{n} \} 
= \{ n_1, n_2, n_3 \ldots \}$, 
where $n_a$ is the number of particles in orbital $a$. Furthermore, let $N_a = 2j_a +1$ be the maximum 
occupation of orbital $a$.
The subspace dimension is then
\begin{equation}
    d_\alpha = \prod_a \left ( \begin{array}{c} n_a \\ N_a  \end{array} \right )= \prod_a \frac{ N_a!}{n_a! (N_a - n_a)!}.
\end{equation}
Note that this dimension includes all orientations, that is, all values of total $J_z$ or $M$.
In typical configuration-interaction calculations one fixes $M$;  the inclusion of all values of $M$, however, 
makes for much simpler formulas. (This also means the trace weights more heavily 
states with higher total $J$, which is a price one must pay for simplicity.)

To compute the centroids, 
define the monopole potential
\begin{eqnarray}
    U(ab) & = & \frac{ \sum_J (2J+1) V_J(ab,ab)}{\sum_J (2J+1)}\\
    \nonumber
 & = &    \frac{1+ \delta_{ab}}{N_a(N_b - \delta_{ab})}\sum_J (2J+1) V_J(ab,ab).
 \label{monopot}
\end{eqnarray}
Then the configuration centroid is 
\begin{equation}
    E_\alpha = \sum_a n_a \epsilon_a + \frac{1}{2} \sum_{ab} n_a (n_b - \delta_{ab}) U(ab).
\end{equation}
This very simple formula enables near instant calculation of traces/centroids even of subspaces 
with enormous dimensions. 

\section{Many-body truncations}
\label{sec:truncate}

Naively, if one has $N_p$ fermions in $N_s$ single-particle states, there are 
\begin{equation}
    \left ( 
    \begin{array}{c} N_s \\ N_p \end{array}
    \right ) = \frac{ N_s!}{ N_p! (N_s - N_p)!}
\end{equation}
possible configurations, which grows exponentially. In practice, fixing quantum 
numbers such as parity or total $J_z = M$ reduces the overall number, but the explosive growth remains. Thus truncation schemes for configuration-interaction have long been 
a topic of investigation. 

One way to truncate  is through group-theory inspired subspaces, 
such as seniority truncation ~\cite{PhysRevC.59.2033} or SU(3) or Sp(3,R). 
Far simpler, however, are truncations based upon orbital occupation. 
To do this, one assigns  an integer weight $w_a$ to each orbital. For example, in the standard approach to the no-core shell model~\cite{barrett2013ab}, the weight is the principal quantum number, $N$, of each oscillator orbital. For particle-hole excitations, 
one might assign a weight of zero to orbitals below an imagined Fermi surface, and a weight of one to orbitals above it. 

Now supposed the many-body basis states are occupation-representation Slater determinants. Each basis state can be conveniently represented 
by a binary word, where a 1 is an occupied single-particle state, and a 0 is an unoccupied single-particle state. (Very convenient for digital computers.) Then the  occupation of an  orbital $a$ will be some 
integer $n_a$. 
Finally, for some basis Slater determinant, the total weight is simply the sum of the weights of 
each occupied single-particle state, that is,
\begin{equation}
    W_i = \sum_a w_a n_a.
\end{equation}
This integer weight acts as a kind of quantum number, and it is relatively straightforward to devise an algorithm which allows only states up to a defined maximum weight.   With a skillful choice of weights, however, one can construct more general truncations, as discussed next.

\subsection{Approximately cutoff-energy dependent (ACE) truncation}

A physically motivated scheme is to truncate the many-body space is one based upon the subspace centroid, $E_\alpha$~\cite{PhysRevC.50.R2274}. High efficiency shell-model codes, however, typically utilize an on-the-fly reconstruction of the Hamiltonian that relies upon abelian quantum numbers (that is, quantum numbers that simply add, such as $j_z$ or multiply, such as parity)~\cite{BIGSTICK}. Here we combine those two ideas: choose the integer orbital weights such that one approximately respects a cutoff in centroid energy, $E_\mathrm{cut}$

In order to optimize the weights and thus a truncated space, one must compute the centroids in an enveloping space. 
Ideally this would be the full configuration space, but that may not always be practical. 

We let ${\cal F}$ label the enveloping space, so that $\alpha \in {\cal F}$ defines the set
of configuration subspaces in the enveloping space.  Recall that each partition is defined by 
a set of occupations $\vec{n} = \{ n_a \}$. For a given set of orbital weights, 
$\vec{w} = \{ w_a \}$, the total weight of a partition is 
\begin{equation}
    W_\alpha = \sum_a n_a w_a.
\end{equation}
For a selected maximum $W_\mathrm{max}$, a truncation ${\cal P}(\vec{w}, W_\mathrm{max}$ is defined by all $\alpha$ with $W_\alpha \leq W_\mathrm{max}$. (Note that in {\tt BIGSTICK}, 
this definition is exact. In {\tt tracer}, however, one gets an approximate truncation 
where one draws from all $\alpha \in {\cal F}$. In most cases the two spaces overlap, if ${\cal F}$ is defined sufficiently generous.)

We introduce a fitness function, $\Phi$, which we want to minimize. Specifically:

\begin{itemize}

\item We want to add a reward (a lowering of $\Phi$) for configurations with centroid energy $E_\alpha \leq E_\mathrm{cut}$ 
    and which \textit{are} in the truncation ${\cal P}$, that is, have $W_\alpha > W_\mathrm{max}$;


    \item We also want to add a penalty for configurations which \textit{are}  in the truncation ${\cal P}$, that is, have $W_\alpha \leq W_\mathrm{max}$, but with centroid energy above the cut $E_\alpha >  E_\mathrm{cut}$;

    \item  We want partitions with a larger number of states  (larger $d_\alpha$) to have  greater weights;

    \item  We want penalties/rewards to be larger penalty for deeper configuration, that is, 
    for which $E_\mathrm{cut} - E_\alpha$ is larger.

\end{itemize}
We can combine these into a simple formula 
\begin{equation}
    \Phi(E_\mathrm{cut},\vec{w},W_\mathrm{max},{\cal F}) = \sum_{\alpha \in {\cal P}(\vec{w}, W_\mathrm{max})} d_\alpha (E_\alpha - E_\mathrm{cut} ).
\end{equation}
 All configuration in ${\cal P}$ are accounted for: if $E_\alpha \leq E_\mathrm{cut}$, as
a reward,  or $E_\alpha > E_\mathrm{cut}$, a penalty; otherwise one could accidentally minimize $\Phi$ by leaving 
out all configurations! 
We found this prescription for a fitness function works well.

Here is how the minimization process work. Starting from some initial set of orbital weights, 
the code generates all the configurations within the enveloping space ${\cal F}$.
The configurations are stored as occupations $\vec{n}$; thus, given some orbital weight 
scheme $\vec{w}$, it is easy to quickly generate all the configuration weights $W_\alpha$. 
In the process of minimization computing the $\{ W_\alpha \}$ is typically done thousands or tens of thousands of times.

For any fixed $\vec{w}$, the code scans over $W_\mathrm{max}$ to minimize the fitness function. 
With that in mind, the code does a Monte Carlo search over $\vec{w}$ to find a weight scheme 
(and corresponding $W_\mathrm{max}$) that minimizes the fitness function. 

Finally, to more fully explore the fitness landscape, the code will at intervals 
reset to a random starting point. Experience shows that resetting after about 200 steps works well.

\section{Compiling and running the code}\label{running}
\label{sec:run}

The code compilation is straightforward. The default compiler is the 
widely available GNU {\tt gfortran} compiler, but one can easily edit 
{\tt makefile} to use another compiler. No special libraries are called. 
Although we have experimented with OpenMP parallelization, in nearly all 
cases the code is sufficiently fast to not require parallelization. 

To compile:
\begin{verbatim}
make tracer    
\end{verbatim}
produces the executable {\tt tracer}. Note also that
\begin{verbatim}
make clean    
\end{verbatim}
will delete intermediate object files, mod (module) files, and the executable, 
for a fresh compile.

To run the code, you need two input files: a file with extension {\tt .sps} 
to define the single-particle space (but this is optional if one is 
working in a no-core shell model space), and one or more interaction files, 
with extension {\tt .int}. The file format is defined in \ref{sec:format}, 
but are exactly the same formats as used in the {\tt BIGSTICK} code.

When you start {\tt tracer}, you get:
\begin{verbatim}
 TRACER 
  Version 10 (July 2025) 
\end{verbatim}
You are asked the following questions:
\begin{verbatim}
  Do you want to write results to a file (y/n)?
\end{verbatim}
and
\begin{verbatim}  
  Do you want to write out configurations 
and configuration centroids ?
\end{verbatim}
In nearly all cases you will want to answer `y' (yes). 
\begin{verbatim}
Enter output name of file (do not include suffix) 
\end{verbatim}
Much as in the {\tt BIGSTICK} code, you are asked for an input file 
which defines the single-particle space (see appendix, or the {\tt BIGSTICK} manual).
\begin{verbatim}
  Enter file with s.p. orbit information (.sps)
  (Enter "auto" to autofill s.p. orbit info )     
\end{verbatim}
If you want to truncate the many-body space, the single-particle space 
will need different weights assigned to each orbital. 
If truncation is possible, you are asked:
\begin{verbatim}
  Would you like to truncate ? (y/n)
  Note: must choose y in order to initiate 
optimization of weights)
\end{verbatim}
Here `y' will mean asking for truncation, `n' means all configurations 
included. If you choose `y' then
\begin{verbatim}
  Max excite =           12
  Min, max excite you allow 
  (This is like Nmax/ Nhw )
  (Note: must include BOTH min AND max )    
\end{verbatim}
Note that {\tt BIGSTICK} only asks for the maximum excitations. 
Here, we allow for a range of excitation, to better allow to separate out 
the contributions from specific excitations.

By entering in a minimum and maximum excitation, even if the choice 
includes all possible excitations, you now have the option to 
optimize the weights
\begin{verbatim}
Would you like to optimize wmax and 
weights for the truncation ? (y/n)    
\end{verbatim}

Conversely, if all orbitals 
are assigned the same weight, you get:
\begin{verbatim}
  using same weights for protons and neutrons 
  Expecting iso format: same weights for p, n 
  Enter # of protons, neutrons 
 ( max =           20 ,          20 )    
\end{verbatim}
Next, you can enter in multiple interaction files 
in {\tt BIGSTICK}-compatible formats, entering {\tt END} to
conclude
\begin{verbatim}
Enter interaction file name (.smint/.int)
Enter "xpn" or "upn" to select explicit 
proton-neutron input format 
(Note: "xpn" is preferred, normalized convention, 
but use "upn" for NuShell-originated files )
  (Enter END to stop or skip ) 
\end{verbatim}
For more details, see the Appendix.

If you do not truncate or do not choose to optimize the weights, 
{\tt tracer} will compute the centroids. To the screen is written 
(this is the case for $^{60}$Fe in the $fp$ shell using the GX1A interaction)
\begin{verbatim}
 1  there are           60  configurations 
 2  there are           60  configurations 
\end{verbatim}
(1 = protons, 2 = neutrons),
\begin{verbatim}
  Monopole potentials written to upot.dat
  expect         3600  configurations 
  COMBINED CONFIGS 
        3600  configurations
           1502182564  states in total 
  Centroid =   -173.92364600362626     
  
  Lowest configuration has energy   -219.141     
  proton configuration 1 , neutron configuration 3
  
  width of centroids =    9.421   
  3rd moment of centroids =   -8.258E-002
  4th moment of centroids =    2.909     
\end{verbatim}
The 3rd and 4th moments are scaled by the width (square root of second moment). 
In most cases the units are MeV, but depend upon the input file.

Fig.~\ref{fig:fe60} plots an example of the energy centroids for $^{60}$Fe.




\subsection{Example results: optimization of ACE truncations}

\begin{figure}[h]
    \centering
    \includegraphics[width=0.5\textwidth]{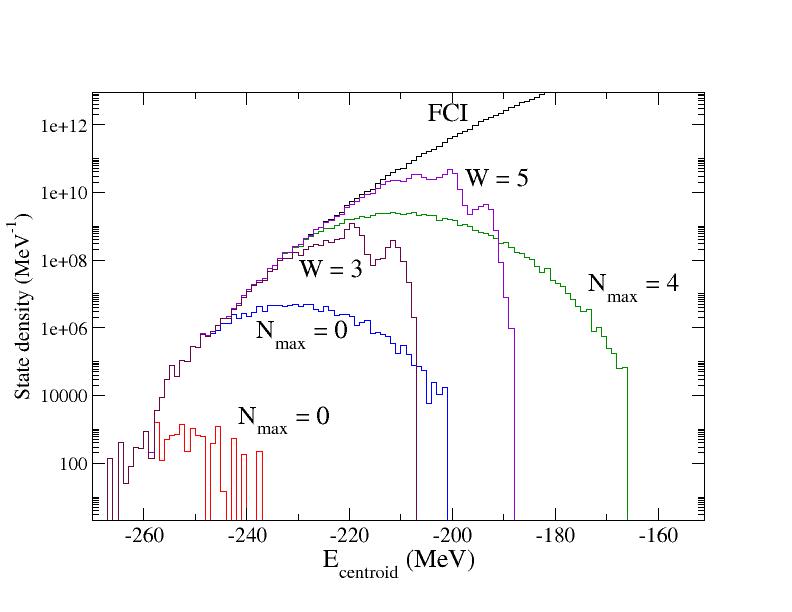}
    \caption{Distribution of number of states (not levels) in configurations versus configuration centroids for $^{40}$Ar computed in the $sd$-$pf$ space, for different orbital weights and truncations. FCI is the full space. The $N_\mathrm{max}$ denotes the number of particles out of the $sd$ shell into the $pf$ shell, while $W$ is the relative excitation using an optimized set of orbital weights. Data binned in 1-MeV bins.
    }
    \label{fig:ar40}
\end{figure}

As an example, we take the case of $^{40}$Ar in the \textit{sdpf} cross-shell space, shown 
in Fig.~\ref{fig:ar40}, using the sdpfmu-db interaction. We start with the standard 
weighting, where $w=2$ for the $sd$ shell and $3$ for the $pf$ shell; hence, $N_\mathrm{max}$ 
denotes the number of particles excited out of the $sd$ shell into the $pf$ shell. 
The full space (positive parity, $M+0$ has a basis dimension of over $9 \times 10^{14}$, 
while $N+\mathrm{max}=0$ has a dimension of only 1566, $N_\mathrm{max}=2$ a dimension of 
9 million, and $N_\mathrm{max}=4$ a dimension of 4.6 billion.

If you choose to optimize, you will be asked
\begin{verbatim}
Enter the number of total iterations you would like 
the program to run for the optimization     
\end{verbatim}
While it usually only takes about 200 iterations for an optimization to complete, it is 
recommended to run multiple cases with different random seeds, about 5 to 10 cases.
Hence one answers a total of 1000 to 200 iterations, and when asked
\begin{verbatim}
  Enter how often to restart (zero to not restart) 
  (Typical value ~200)    
\end{verbatim}
a value of 200 is reasonable.  However for large cases this can take time.

Next,
\begin{verbatim}
Include the real number energy cutoff
for your truncation     
\end{verbatim}
For the $^{40}$Ar case above, we choose a cutoff energy of -200 MeV. 

As the code runs, you will get intermediate results, including the results on
the cost function for the current set of weights
\begin{verbatim}
  Iteration:  160 , Initial cost:  8151709768804, 
                     Lowest cost: 7907824875702    
  New weights:   1   4   2   12   4    5   2
 
  Optimal w_max:           16
  Trial Cost:    8699512636445.4961     
  Prior Cost:    8676206001372.2734     
  Difference:    23306635073.222656     
  Percentage:   0.26862703662794957      %
\end{verbatim}
After all the iterations have been carried out, the winning set of weights is displayed:
\begin{verbatim}
 Final Iteration: 1999 ; Energy cut-off:   -200.000   
  Optimal Max Excitation:           5
  Optimal weights:  1   2    1   3   2   3   2
  Final cost:                       5295343403008     
  Initial cost:                     8151709768804       
\end{verbatim}
This can also be written to file:
\begin{verbatim}
Enter name out of output .sps file 
Enter "none" to not write to file 
DO NOT USE THE SAME NAME AS ORIGINAL .sps FILE    
\end{verbatim}
The new {\tt .sps} file with optimized weights thus looks like
\begin{verbatim}
iso
    7
  1.0  0.0  0.5   1
  0.0  2.0  1.5   2
  0.0  2.0  2.5   1
  1.0  1.0  0.5   3
  1.0  1.0  1.5   2
  0.0  3.0  2.5   3
  0.0  3.0  3.5   2    
\end{verbatim}

In Fig.~\ref{fig:ar40}, we plot the $W=3$ case, with an $M=0$ dimension of 604 million, and 
the $W=5$ case, with a dimension of 35 billion. Although not perfect, one can see 
that this truncation captures low-energy configurations missed by the standard $N_\mathrm{max}$ 
truncation.  Of course, this truncation does not allow exact separation of spurious center-of-mass motion, but neither does the original truncation.  Such spurious behavior can be suppressed 
using the Lawson-Palumbo presscription.

One can of course play around with cutoffs.  In our experience, modest changes in the cutoff does not require re-optimization, but running tracer to check is a good idea.



\section{Conclusion}

We have presented the code {\tt tracer}, available for download from {\tt github.com/cwjsdsu/tracer}, 
which efficiently computes the configuration centroids for shell-model interactions. The code can also optimize the orbital weights for an approximation truncation in energy.  Future work will 
investigate the efficacy of these optimized weights.

\section{Acknowledgements}

This material is based upon work supported by the U.S.~Department of Energy, Office of Science, Office of Nuclear Physics, under Award Number DE-FG02-03ER41272. The code is free and open-source, released under the MIT License. 

\bibliography{johnsonmaster}

\appendix

\section{Format of input files}
\label{sec:format}

\subsection{Single-particle space}

The single-particle space is defined one of two ways. Either  read in a file defining the single-particle space, or, for so-called \textit{no-core 
shell model} calculations, automatically generate the basis in a pre-defined form, using the autofill or `{\tt auto}' option.

 For consistency, we generally refer to 
\textit{orbitals} as single-particle spaces labeled by angular momentum $j$ but not 
$j_z$, while \textit{states} are labeled by both $j$ and $j_z$. 

Our default format for defining the single-particle space are derived from the format for {\tt OXBASH/NuShell/NuShellX} 
files.  A typical file is the {\tt sd.sps} file:
\begin{verbatim}
! sd-shell
iso
3
       0.0  2.0  1.5  2
       0.0  2.0  2.5  2 
       1.0  0.0  0.5  2
\end{verbatim}
There is no particular formatting (spacing) to this file.  Any header lines starting with an exclamation point ! or a hash mark $\#$ are skipped over. 
The first non-header line denotes about the isospin symmetry or lack thereof. 
{\tt iso} denotes the single-particle space for both species is the same; one can still 
read in isospin breaking interactions. 
The second line (3 in the example above) is the number of single-particle orbits. 
The quantum numbers for the single-particle orbits as listed are: $n, l, j, w$; the first three numbers 
are real or integers, $j$ is a real number.
 $n$ is the radial quantum number, which play no role except to 
distinguish between different states. $l$ is the orbit angular momentum and $j$ is 
the total angular momentum; for the case of nucleons $j = l \pm 1/2$. 
While $n$ is not internally significant, it aids the human-readability 
of the {\tt .sps} files; in addition, it can be invaluable as input to other code 
computing desired matrix elements.

The last `quantum number,' $w$, is the \textit{weight} factor, used for many-body 
truncations,  described  in Section \ref{sec:truncate}. It must be a nonnegative 
integer. 

{\tt tracer} can handle any set of single-particle orbits; the only requirement is that 
each one have a unique set of $n,l,j$.
For example, one could have a set of 
$l= 0, j = 1/2$ states:
\begin{verbatim}
iso
4
       0  0  0.5  0
       1  0  0.5  0
       2  0  0.5  0
       3  0  0.5  0
\end{verbatim}

As of the current version, one cannot define completely independent 
proton and neutron spaces.  One can however specify two variations where protons and neutrons can have different weights.
The preferred format is {\tt pnw}, where one lists the quantum numbers as well as the proton and neutron weights in two columns:
\begin{verbatim}
pnw
3 
       0.0  2.0  1.5  3   3
       0.0  2.0  2.5  2   3
       1.0  0.0  0.5  2   3
\end{verbatim}
For some more details on using the {\tt pnw} format, especially in delineating different proton and neutron valence spaces, see Section~\ref{protonneutron} below.

The ordering of the single particle orbits is important and must be consistent with 
the input interaction files.  If one uses our default-format interaction files, one 
must supply a {\tt .sps} file.

For no-core calculations, where a standard ordering of orbital exists, one can use the \textit{autofill} option  for defining the single-particle states, by entering {\tt auto} in place of the name of the {\tt .sps} file:
\begin{verbatim}
  Enter file with s.p. orbit information (.sps)
  (Enter "auto" to autofill s.p. orbit info ) 
auto
  Enter maximum principle quantum number N 
  (starting with 0s = 0, 0p = 1, 1s0d = 2, etc. ) 
\end{verbatim}
The autofill option creates a set of single-particle orbits assuming a harmonic oscillator, 
in the following order: $0s_{1/2},$ $0p_{1/2},$$0p_{3/2},$$1s_{1/2}, 0d_{3/2}, 0d_{5/2}$, etc.,  that is, for given $N$, in 
order of increasing $j$, 
up to the maximal value $N$.
 It also associates a value $w$ equal to the principal 
quantum number of that orbit, e.g., $2n+l$, so that $N$ above is the maximal principal 
quantum number.   So, for example, if one choose the principle quantum number $N= 5$ this includes up to the 
$2p$-$1f$-$0h$ shells, which will looks like
\begin{verbatim}
iso
21
0.0  0.0  0.5  0
0.0  1.0  0.5  1
0.0  1.0  1.5  1
1.0  0.0  0.5  2
0.0  2.0  1.5  2
0.0  2.0  2.5  2
...
1.0  3.0  3.5  5
0.0  5.0  4.5  5
0.0  5.0  5.5  5
\end{verbatim}

All truncation is based upon the $w$ weight factors. In most applications, both protons and neutron orbits have the same 
weights, and one typically truncates equally.  A more general truncation scheme is possible.

\subsection{How to handle `different' proton-neutron spaces}

\label{protonneutron}

As of the current version, neither {\tt BIGSTICK} nor {\tt tracer} can directly handle independently-defined proton and neutron spaces. You can, however trick it into behaving that way, with a 
small cost.   Both involve deft usage of the truncation and, in many cases, of particle-hole truncation.

Let's consider two toy cases.  First, suppose the proton and neutron spaces are entirely separate. For example, let's suppose valence  
protons occupy only the $0f_{7/2}$ space and valence neutrons only the $1p_{3/2}$. The {\tt .sps} file can look like:
\begin{verbatim}
iso
   2
0.0  3.0  3.5  0
1.0  1.0  1.5  1
\end{verbatim}
By choosing a {\tt Max excite} of zero, you will assure no particles are excited out of the $0f_{7/2}$ into the $1p_{3/2}$. (It is your responsiblity to 
set up the correct interaction file. You do not have to include cross-shell matrix elements if they are not needed; however if they are included, 
they will induce an effective single-particle energy so choose wisely.)

A more general, and \textbf{recommended}, approach is to use the {\tt pnw} format: suppose you want protons active in $0f_{7/2}$, $1p_{3/2}$ and $1p_{1/2}$, and neutrons in 
$1p_{3/2}$, $1p_{1/2}$, and $0f_{5/2}$.  Set up the {.sps} file
\begin{verbatim}
pnw
4  
       0.0  3.0  3.5  0    0
       1.0  1.0  1.5  0   99
       1.0  1.0  0.5  0   99
       0.0  3.0  2.5 99   99
\end{verbatim}
It is required that the proton and neutron orbits be the same, though the weight factors $w$ is the last column can differ. 
A weight of 99 signals that the orbital is `sterile' for either protons or neutrons, which means it will not be used.
Again, choosing {\tt Max excite} of zero will keep the protons and neutrons in their respective valence spaces.  If the valence spaces are 
significantly different, we strongly recommend utilizing particle-hole conjugation for the neutrons.  

One can make the truncations even more complex, for example allow a few protons to be excited but no neutrons, by careful usage of the 
options provided.   For example, setting
\begin{verbatim}
pnw
4  
       0.0  3.0  3.5  0     99
       1.0  1.0  1.5  1     99
       1.0  1.0  0.5  1     99
       0.0  3.0  2.5  99     0
\end{verbatim}
and setting the maximum truncation to 2, you can excite up to 2 protons out of the $0f_{7/2}$ into the 
$1p_{3/2}$ and $1p_{1/2}$ orbits, but none into the $0f_{5/2}$, while you will have only neutrons 
in the $0f_{5/2}$ but none in the $0f_{7/2}$-$1p_{3/2}$ -$1p_{1/2}$ orbits.

Here you must carefully consider the nature of the proton-neutron interaction.  Suppose you wanted 
four valence protons in the $0f_{7/2}$ -$1p_{3/2}$ - $1p_{1/2}$ space and 2 neutrons in the 
$0f_{5/2}$.  You could also set 
\begin{verbatim}
pnw
4  
       0.0  3.0  3.5  0    0
       1.0  1.0  1.5  1    0
       1.0  1.0  0.5  1    0
       0.0  3.0  2.5  99  99
\end{verbatim}
Because the $0f_{7/2}$ -$1p_{3/2}$ - $1p_{1/2}$ space has a total of 14 states, you have have instead 
set valence $N=14+2=16$.  With {\tt max excite} = 2, the neutrons in the $0f_{7/2}$ -$1p_{3/2}$ - $1p_{1/2}$ space
will be fixed. 

In this example, while the neutrons in $0f_{7/2}$ -$1p_{3/2}$ - $1p_{1/2}$ are fixed, they can have matrix 
elements with other particles, producing a change in single-particle energies. You should therefore understand 
carefully both your model space and your interactions. 

\textbf{Important}: Be careful in how you read in your interaction file. Although you are treating the proton and neutron spaces separately, in many cases the supplied interaction file, at least for empirical valence spaces, 
will still be in {\tt iso} format (see the next section for detail). You can test this by trying a small cases in your space, for example, just two protons and two neutrons. If you have set up correctly, you will get integer values of $J$. Alternately, if you get irrational values of $J$, the most likely culprit is that you have put in the wrong format for the interaction file.  

\subsection{Interaction files}

After the model space is defined, {\tt tracer}, like {\tt BIGSTICK}, needs interaction matrix elements. All matrix elements are defined in the one- and two-body-space.  

The default format for two-body interaction file is derived from {\tt OXBASH/NuShell} and always ends in the extension {\tt .int}.  When entering the 
name of the file, only enter the name, not the extension, i.e., {\tt usdb} not {\tt usdb.int}. 
\begin{verbatim}
!  Brown-Richter USDB interaction
63     2.1117   -3.9257   -3.2079  
  2  2  2  2    1  0  -1.3796
  2  2  2  1    1  0   3.4987
  2  2  1  1    1  0   1.6647
  2  2  1  3    1  0   0.0272
  2  2  3  3    1  0  -0.5344
  2  1  2  1    1  0  -6.0099
  2  1  1  1    1  0   0.1922
  2  1  1  3    1  0   1.6231
  2  1  3  3    1  0   2.0226
  1  1  1  1    1  0  -1.6582
  1  1  1  3    1  0  -0.8493
  1  1  3  3    1  0   0.1574
. . . 
\end{verbatim}
There is no specific spacing for this file. {\tt tracer} will skip any header lines starting with ! or $\#$.
The first line is

\smallskip

\textit{ Ntbme \qquad   spe(1) \qquad  spe(2) \qquad   spe(3) ... }

\smallskip

where \textit{Ntbme} is the number of \textit{two-body matrix elements} (TBMEs)
 in the rest of the 
file, and \textit{spe(i)} is the \textit{single-particle energy} of the $i$th orbit.  
(Note: older version required only 10 single particle energies are on each line. This has been changed and is no longer 
required.)

The rest of the file are the two-body matrix elements. This is defined as 
\begin{equation}
V_{JT}(ab,cd) = \langle ab; JT | V | cd; JT \rangle,
\label{tbmedef}
\end{equation}
where $a,b,c,d$ label orbits, as ordered in the {\tt .sps} file or as created by the autofill option; 
$J$ and $T$ are the total angular momentum and total isospin of the two-body states 
$| ab; JT \rangle$, which are normalized. This follows the convention of Brussaard and
Glaudemans.  Each matrix element is read in as

\smallskip

\textit{ a \qquad b \qquad  c \qquad d \qquad J \quad T } \qquad $V_{JT}(ab,cd)$ 

\smallskip

For input purposes, the order of $a,b,c,d$ is not important (as long as one has the 
correct phase), nor is the ordering of the TBMEs themselves.  When reading in the file, 
{\tt tracer} automatically stores the matrix element according to internal protocols, 
appropriately taking care of any relevant phases. 

Matrix elements that are zero can be left out, as long as $Ntbme$ correctly gives the 
number of TBMEs in the file. \textbf{More than one file can be read in; enter} {\tt end} 
\textbf{to  tell} {\tt tracer} \textbf{you are finished reading interaction files. }

\subsection{Scaling and autoscaling}

\label{scaling} 

Empirical studies with phenomenological interactions have found best agreement with experiment if one scales the two-body matrix elements 
with mass number $A$. (There is some justification based upon the scaling of harmonic oscillator wave functions with $A$). A standard scaling factor 
is 
\begin{equation}
\left ( \frac{ A_\mathrm{0} }{A} \right )^x
\end{equation}
where $A_\mathrm{0}$ is the reference mass number (typically $A$ of the frozen core +2, as it is fit to the interaction of two particles above
the frozen core), $A$ is the mass of the desired nucleus, and $x$ is an exponent, typically 
around $1/3$.  To accomodate this scaling, when reading in the default format, {\tt tracer} requests
\begin{verbatim}
  Enter scaling: spescale, A0,A,X 
 ( spescale scales single particle energies, 
  while TBMEs are scaled by (A0/A)^X ) for TBMEs 
  (If A or X = 0, then TBMES scaled by A0 ) 
\end{verbatim}
Typically the single particle energies are unscaled, but we allow for it. A typical entry, for example for the USDA/B interactions (\citep{PhysRevC.74.034315}), would 
be 
\begin{verbatim}
1   18   24   0.3
\end{verbatim}
Here the single particle energies are unscaled, the core has mass number 16 and hence the reference mass $A_0$ is 18, the target mass in this case has mass number $A=$ 24, and the 
exponent is 0.3.    Whoever provides the interaction has to provide the exponent.  If unsure, just enter
\begin{verbatim}
1, 1, 1, 1
\end{verbatim}

Many files used with {\tt NuShell} have autoscaling. For example, for the USDA/B file, the first lines are 
\begin{verbatim}
-63  1.9798   -3.9436   -3.0612  16.0  18.0 0.3
\end{verbatim}
 A negative integer for the number of two-body matrix elements  (here, -63) initiates autoscaling.  
 The next three numbers are the single-particle energies, and the next numbers are $A_\mathrm{core}$, the reference mass, 
 and the exponent. If {\tt BIGSTICK} encounters a negative integer for the number of two-body matrix elements, it autoscale the 
 two-body matrix elements as described above. To turn off autoscaling, change -63 to 63.
 
 Keep in mind that not all interactions will be scaled.  \textit{Ab initio} interactions are almost never scaled, and `phenomenological'
 interactions depend on how they were derived and fit. See your interaction provider for more information.
 
 If you enable autoscaling (by setting  the number of matrix elements negative) and set the three parameters ($A_\mathrm{core},$ reference mass, and exponent) to zero, 
 i.e., so it looks like
 \begin{verbatim}
 -63     1.9798   -3.9436   -3.0612 0 0  0
 \end{verbatim} then all parameters will be left unchanged, that is, autoscaled by one; furthermore,  you will not be asked to enter in scaling factors. Autoscaling in both forms may be useful for impatient users and and users not comfortable with scaling.

\subsection{Proton-neutron and other isospin-breaking formats}

\label{pnham}

Often one needs to break isospin. {\tt tracer} using proton-neutron formatted inputs for 
breaking isospin. 
We recommend the explicit proton-neutron formalism.  Here one has separate labels for proton and neutron 
orbits; however, at this time \textbf{the proton and neutron orbits must have the same quantum numbers and be listed in the same order.}
For example, one might label the proton orbits $1= 0d_{3/2}$, $2=0d_{5/2}$, and $3=1s_{1/2}$. Then the neutron orbits must be 
$4= 0d_{3/2}$, $5=0d_{5/2}$, and $6=1s_{1/2}$.

While {\tt tracer} generally allows for arbitrary order, for the proton-neutron matrix elements the proton labels must be in the first and third columns and neutron labels in the second and fourth columns, 
that is, for $V_J(ab,cd)$, $a$ and $c$ \textit{must} be proton labels and $b,d$ must be neutron labels. 
With twice as many defined orbits, one must also provide separate 
proton and neutron single particle energies.  As an example, here is part of the file of the $p$-shell Cohen-Kurath matrix elements with good isospin:
\begin{verbatim}
! ORDER IS:  1 = 1P1/2    2 = 1P3/2 
   15      2.419      1.129
  1   1   1   1        0   1      0.2440000
  1   1   1   1        1   0     -4.2921500
  2   1   1   1        1   0      1.2047000
  2   1   2   1        1   0     -6.5627000
  2   1   2   1        1   1      0.7344000
  2   1   2   1        2   0     -4.0579000
  2   1   2   1        2   1     -1.1443000
  2   2   1   1        0   1     -5.0526000
\end{verbatim}
and here is an excerpt in proton-neutron formalism
\begin{verbatim}
    34    2.4190    1.1290    2.4190    1.1290
   1   3   1   3     0   1   0.24400
   1   1   1   1     0   1   0.24400
   3   3   3   3     0   1   0.24400
   1   3   1   3     1   1  -4.29215
   1   3   1   4     1   1  -0.85185
   1   3   2   3     1   1   0.85185
   1   3   2   4     0   1  -5.05260
   1   1   2   2     0   1  -5.05260
   3   3   4   4     0   1  -5.05260
   1   3   2   4     1   1   1.76980
   1   4   1   4     1   1  -2.91415
   1   2   1   2     1   1   0.73440
   3   4   3   4     1   1   0.73440
\end{verbatim}
In no case are headers required, but they do help as a check for the definition of the orbits. 
In the explicit proton neutron format $T$ is given in the sixth column but not actually used. 

\bigskip

There is one more question of convention one must deal with: the normalization of the two-body states in the definition of 
matrix elements.  All formats assume two-proton and two-neutron states are normalized, and states with good isospin are normalized. 
Files set up for {\tt NuShellX}, however, have \textit{unnormalized} proton-neutron states.  

{\tt tracer} can read in default-format proton-neutron interactions with either normalized (`{\tt xpn}' or explicit proton-neutron) 
or unnormalized (`{\tt upn}' or unnormalized proton-neutron) conventions. In both cases the files also include proton-proton and neutron-neutron
matrix elements, with normalized states. 

The relationship between the two is
\begin{equation}
V_J^{xpn}(a_\pi b_\nu, c_\pi d_\nu) = \frac{\sqrt{(1+ \delta_{ab})(1+ \delta_{cd})}}{2} V_J^{upn}(a_\pi b_\nu, c_\pi d_\nu) 
\label{upn2xpn}
\end{equation}
Here we have marked the orbits $a,c$ as proton and $b,d$ as neutron, but the Kronecker-$\delta$s refer only to the quantum numbers $n,l, j$. 
For example, in the $sd$ shell, with the labels mentioned above, 
$$
V_J^{xpn}(16,25) = \frac{1}{\sqrt{2}} V_J^{upn}(16,25)
$$
because proton orbit 1 $(0d_{3/2})$ and neutron orbit 6 $(1s_{1/2})$are different, but proton orbit 2 and neutron orbit 5 are both $d_{5/2}$

It is up the user to know whether or not the file uses normalized or unnormalized proton-neutron states. If the file was originally produced 
for use with {\tt NuShellX}, it is almost certainly the latter. 

(This arises out of the conversion of normalized isospin wave function to normalized proton-neutron wave functions and the result matrix elements.
One finds
\begin{equation}
V_J^\mathrm{pn}(ab,cd) = \frac{ \sqrt{(1+\delta_{ab})(1+\delta_{cd})}}{2} \left [ V_{J,T=0}^\mathrm{iso} (ab,cd) + 
V_{J,T=1}^\mathrm{iso} (ab,cd) \right ],
\label{xpndef}
\end{equation}
but the unnormalized convention yields the simpler
\begin{equation}
V_J^\mathrm{upn}(ab,cd) = V_{J,T=0}^\mathrm{iso} (ab,cd) + 
V_{J,T=1}^\mathrm{iso} (ab,cd). \label{upndef}
\end{equation}
While our preference is for the former, given the prominence  of the latter  through {\tt NuShellX} we include it as an option.)

\bigskip

Because of this, {\tt tracer} will ask if the interaction is normalized.

As with default-format isospin-conserving files, the file name must be {\tt xxxx.int}, but 
the user enters in just `{\tt xxxx}'. 

Also as with default-format isospin-conserving files, after entering the name of the file, the user 
is prompted for scaling. For maximal flexibility, there are two layers of possible scaling. The first 
is the standard phenomenological scaling:
\begin{verbatim}
Enter global scaling for spes, A,B,X ( (A/B)^X ) 
for TBMEs (If B or X = 0, then scale by A ) 
1 18. 24. 0.3
\end{verbatim}
These scalings are applied to all single particle energies and to all two-body matrix  elements.


One can mix all of these different formats. You can read in an isospin-conserving file, a proton-neutron format file, and so on, in
any order. To stop reading in interaction files, enter `{\tt end}' at the prompt.

\section{Format of output files}

There are five useful output files. (A sixth output file, with extension 
{\tt .cmtmp}, is only used as a temporary file for optimization)

\subsection{Monopole potential}

The elements of the monopole potential, $U(ab)$ given in Eq.~(\ref{monopot}), 
are written to the generic files {\tt upot.dat}. While most users will not 
need this file, it can be combined with orbital occupation information 
from {\tt BIGSTICK} to compute effective single-particle energies.

The first line of {\tt upot.dat} is the number of proton and neutron orbitals. It is assumed the orbital 
order is known separately.  The example here is for the GX1A interactionin the $fp$ valence space, for mass $A=60$; the only mass-dependence is any scaling 
of the two-body matrix elements.
Following are single-particle energies
\begin{verbatim}
#proton / neutron spes 
    1  -8.62399960      -8.62399960    
    2  -5.67929983      -5.67929983    
    3  -1.38290000      -1.38290000    
    4  -4.13700008      -4.13700008    
\end{verbatim}
(the first columne is the orbital label, the next two columns 
are proton and neutron single-particle energies, respectively, taken from 
the input interaction files, then
\begin{verbatim}
#proton potentials 
           1           1 -0.143234283    
           2           1   9.27669406E-02
           2           2 -0.233616114    
           3           1   5.31982183E-02
.....
\end{verbatim}
which is $a,b, U(a,b)$ for protons,
then
\begin{verbatim}
#neutron potentials 
           1           1 -0.143234283    
           2           1   9.27669406E-02
           2           2 -0.233616114    
           3           1   5.31982183E-02
...    
\end{verbatim}
for neutrons, and
\begin{verbatim}
#proton-neutron potentials 
           1           1 -0.881983519    
           1           2 -0.595169306    
           1           3 -0.968209565    
           1           4 -0.539692461    
           2           1 -0.595169306    
...    
\end{verbatim}
for the proton-neutron interaction. In this particular case, because of isospin symmetry, the proton-neutron potential happens to be symmetric.

\subsection{Moments}

The output file with extension {\tt .m} saves the same information written 
to the screen, e.g.
\begin{verbatim}
  single-particle file = fp
           6          14
 gx1a                                                   
 Z =    6 N =   14
 \end{verbatim}
 Here the orbitals are written out explicitly, in the 
 order $n$, $l$, $2 \times j$, and truncation weight $w$:
 \begin{verbatim}
  Proton orbits 
    0    3    7    0
    1    1    3    0
    0    3    5    0
    1    1    1    0
  Neutron orbits 
    0    3    7    0
    1    1    3    0
    0    3    5    0
    1    1    1    0
  Parity =            1
  No W truncation 
        3600  configurations
           1502182564  states in total 
  Centroid =   -173.92364600362626     
  width of centroids =    9.421394443     
  3rd moment =   -8.2583145073178141E-002
  4th moment =    2.9094104034939305     
  Lowest configuration has energy   -219.141     
  proton configuration  1 , neutron configuration  3   
\end{verbatim}
Note that \textit{states}, including all possible values of $M=J_z$, are counted, not levels. This greatly simplifies the calculation.

\subsection{Configuration centroids}

The output file with extension {\tt .cm} saves the configuration 
centroids. The header is the same as for the {\tt .m} file; then it 
lists
\begin{verbatim}
           1
   6   0   0   0
           2
   5   1   0   0
           3
   5   0   1   0
           4
   5   0   0   1    
\end{verbatim}
which means: proton configuration 1 has 6 protons in the first orbital 
and none in the others; configuration 2 has 6 protons in the first orbital and 1 in the second; and so on. This is followed by the neutron configurations.
Finally the commbined configurations are given:
\begin{verbatim}
      1      1           420 -217.631882
      1      2           336 -218.193420
      1      3            28 -219.141296
      1      4          2240 -215.678040
      1      5          3360 -215.546677
      1      6           672 -215.801682
      1      7          2520 -214.472214
\end{verbatim}
The first column is the index of the proton configuration previously written, the second the index of the neutron configuration, the third column is the number of states, and then the configuration centroid.

\subsection{Configuretion centroids data}

The output file with extension {\tt .cmdata} saves the smme information 
as the {\tt .cm} file; however the former is more useful for binning/graphing, while the latter is more human-readable.  The {\tt .cmdata} file contains only the centroids, in the same order as in the {\tt .cm} file, and the dimensions, thus:
\begin{verbatim}
  -217.631882                      420
  -218.193420                      336
  -219.141296                       28
  -215.678040                     2240
  -215.546677                     3360
  -215.801682                      672
  -214.472214                     2520
  -213.647949                     6720
  -213.210052                     2520
  -214.014374                      672
  -212.497208                     3360
...    
\end{verbatim}
This is useful as input for the binned data plotted throughout this paper.

\subsection{Configuration occupations}

Finally, the output file with the extension {\tt .occ} lists the 
configuration centroids and number of states, followed by the combined (proton plus neutron) occupations of the orbitals:
\begin{verbatim}
 -217.631882         420
 14  4  2  0
 -218.193420         336
 14  4  1  1
 -219.141296          28
 14  4  0  2
 -215.678040        2240
 14  3  3  0
 -215.546677        3360
 14  3  2  1
...    
\end{verbatim}
The user is free to modify the code, of course, to produce any desired output in any useful format.

\end{document}